\documentclass[prb,twocolumn,showpacs]{revtex4}
\usepackage{graphicx}% Include figure files
\usepackage{dcolumn}% Align table columns on decimal point
\usepackage{bm}% bold math

\newcommand{\pdiffl}[2]{\frac{\partial #1}{\partial #2}}
\newcommand{\dfrac}[2]{\displaystyle\frac{#1}{#2}}
\newcommand{\SIdens}{g/cm$^3$}

\begin{document}

\preprint{Manuscript BH10045}

\title{The equation of state of solid nickel aluminide}
%\date{27 July 2005}
\date{2 August 2005, revised 2 November 2005, 16 February 2006,
   and 30 June 2007
   -- LA-UR-05-6096}

\author{Damian C. Swift}
\email{dswift@lanl.gov}
\homepage{http://public.lanl.gov/dswift}
\affiliation{%
   P-24, Physics Division, Los Alamos National Laboratory,
   MS~E526, Los Alamos, New Mexico 87545, U.S.A.
}

\author{Dennis L. Paisley}
\affiliation{%
   P-24, Physics Division, Los Alamos National Laboratory,
   MS~E526, Los Alamos, New Mexico 87545, U.S.A.
}

\author{Kenneth J. McClellan}
\affiliation{%
   MST-8, Materials Science and Technology Division,
   Los Alamos National Laboratory,
   MS~G770, Los Alamos, New Mexico 87545, U.S.A.
}

\author{Graeme J. Ackland}
\affiliation{%
   Department of Physics, University of Edinburgh,
   Edinburgh, EH9~3JZ, Scotland, U.K.
}

\begin{abstract}
The pressure-volume-temperature equation of state of the intermetallic
compound NiAl was calculated theoretically, and compared with
experimental measurements.
Electron ground states were calculated for NiAl in the CsCl structure,
using {\it ab initio} pseudopotentials and density functional theory (DFT), 
and were used to predict
the cold compression curve and the density of phonon states.
It was desirable to interpolate and smooth the cold compression states;
the Rose form of compression curve was found to reproduce the
{\it ab initio} calculations well in compression but exhibited
significant deviations in expansion.
A thermodynamically-complete equation of state was constructed for NiAl,
which overpredicted the mass density at standard temperature and pressure (STP) 
by 4\%,
fairly typical for predictions based on DFT
A minimally-adjusted
equation of state was constructed by tilting the
cold compression energy-volume relation by $\sim$7\,GPa to reproduce
the observed STP mass density.
Shock waves were induced in crystals of NiAl
by the impact of laser-launched Cu flyers and by launching NiAl flyers
into transparent windows of known properties.
The TRIDENT laser was used to accelerate the flyers, 5\,mm in diameter and
100 to 400\,$\mu$m thick, to speeds between 100 and 600\,m/s.
Point and line-imaging laser Doppler velocimetry 
was used to measure the acceleration
of the flyer and the surface velocity history of the target.
The velocity histories were used to deduce the stress state,
and hence states on the principal Hugoniot and the flow stress.
Flyers and targets were recovered from most experiments.
The effect of elasticity and plastic flow in the sample and window was
assessed.
The ambient isotherm reproduced static compression data very well,
and the predicted Hugoniot was consistent with shock compression data.
\end{abstract}

% mech props of solids; high-p and shock effects in cond matt; EOS for substances
%\keywords{Suggested keywords}%Use showkeys class option for keyword display

\maketitle

% preprint only
%\listoffigures
%\newpage

%\tableofcontents

\section{Introduction}
The equation of state (EOS) relating the pressure, compression, and temperature
of solids is important to understand the structure of rocky planets
and the response of materials for dynamic loading during impacts or
energetic events such as explosions.  The EOS also provides a test of
our knowledge of underlying physics, particularly the states of electrons
under the influence of the ions, described by many-body quantum mechanics.
Accurate and thermodynamically-complete EOS are valuable
as it is extremely difficult to measure the temperature of material in
a shock wave experiment, and temperature is a key parameter determining
the rate at which plastic flow, phase transitions, and chemical reactions
occur.

Several approaches have been devised to calculate the EOS essentially from first
principles.
One way of classifying the approaches is into those in which the electrons
are treated explicitly, and those in which the effect of the electrons is
subsumed into effective interatomic potentials.
Although interatomic potentials can be derived from calculations in which the
electrons are included explicitly \cite{Moriarty95}, it can be difficult to
represent the non-local aspects of the electron wavefunctions faithfully
via an interatomic potential, and additional complications arise in
deriving cross-potentials between dissimilar types of atom, which are
necessary for calculations of compounds and alloys.
With the electrons treated explicitly, some EOS calculations have been
made for elements of low atomic number in which exchange and correlation
in the electron wavefunctions have been treated rigorously using 
quantum Monte-Carlo techniques \cite{Ceperley01},
but these calculations are computationally intensive and 
have spanned relatively narrow ranges in state space.
Almost universally, exchange and correlation between the electron states
have been treated using variants of the Kohn-Sham density functional theory
(DFT) \cite{Hohenberg64,Kohn65,Perdew92,White94},
which allows the electrons to be represented efficiently 
as single-particle states.
Similarly, excitations of the electrons and ions have been represented
with varying rigor.
The electronic heat capacity may be treated with simple approximations
such as the Sommerfeld model \cite{Ashcroft76}, 
by populating the density of electron energy levels 
calculated at zero temperature,
with a density of levels which is
calculated consistently with the excitations \cite{Bennett85,Liberman90},
or it may be ignored altogether in many situations sampling states
from room temperature to an electron-volt or so.
Ionic motion has been incorporated through a Gr\"uneisen model (usually
fitted to the zero-temperature energy-volume relation)
\cite{Dugdale53}, 
by performing simulations of the classical motion of the atoms under the
action of interatomic potentials or forces from the electron states
\cite{Kress01},
and by calculating the normal modes of the crystal lattice and populating them
as phonons.

We have previously predicted EOS and phase diagrams for elements using 
electron ground states calculated using DFT, electron excitations
into the zero-temperature band structure, and phonon modes calculated 
from electronic restoring forces as atoms are displaced from equilibrium
\cite{Swift_sieos_01}.
An attraction of the method is that compounds and alloys can in principle
be treated in exactly the same way as elements.
We have subsequently demonstrated that the electron ground states can be
found, and EOS estimated using a Gr\"uneisen treatment 
of the thermal excitations,
for the stoichiometric alloy NiTi \cite{Swift_NiTiEOS_05}.
A careful study of the {\it ab initio} phonon modes has been made at zero
pressure for the stoichiometric alloy NiAl \cite{Ackland03}.
Here, we report a more rigorous calculation of the EOS for 
NiAl using {\it ab initio} phonons \cite{Ackland02} and electronic excitations,
and comparing with static and shock compression data.
We also investigate the use of an empirical relation for
the zero-temperature isotherm of NiAl.

Shock wave data have been obtained mainly from the impact of flyers
launched by the expansion of compressed gases.
In our experiments, the flyer was accelerated by the expansion of a
confined plasma, heated by a laser pulse.
The interpretation of shock experiments often depends on the properties
of other materials in the assembly, such as transparent windows,
and these may be affected by time-dependent phenomena such as plastic flow.
We discuss the validity of our laser-flyer experiments for measuring
the EOS of NiAl.

\section{Theoretical equation of state}
A thermodynamically complete EOS can be expressed as any thermodynamic
potential with respect to its two natural variables, such as 
specific internal energy $e$ in terms of specific volume $v$ and 
specific entropy $s$. 
The atomic properties of matter lead naturally to expressions for 
contributions to the internal energy
in terms of the volume (or mass density $\rho=1/v$) and temperature $T$.
Following our previous study of the EOS of elements \cite{Swift_sieos_01},
$e$ was notionally split into the cold compression curve $e_c$,
lattice-thermal contribution $e_l$, and electron-thermal contribution $e_e$:
\begin{equation}
e(\rho,T) = e_c(\rho) + e_l(\rho,T) + e_e(\rho,T).
\end{equation}
The relation $e(\rho,T)$ was then used to construct the thermodynamically
complete relation $e(s,v)$ and hence other desired quantities such as
pressure, by invoking the second law of thermodynamics as described later.
The cold compression energy $e_c$ was calculated from the ground state of the
electrons with respect to stationary atoms,
the lattice-thermal energy $e_l$ from the phonon modes of the crystal lattice,
and the electron-thermal energy from the zero-temperature 
electron band structure
though in practice it was a small contribution to the EOS in the solid regime.

\subsection{Electron ground state calculations}
Over the density range of interest, the core electrons 
are affected relatively little by each atom's environment,
so pseudopotentials were used in preference to all-electron calculations.
The electronic ground states were calculated quantum mechanically using
the CASTEP computer code.  This code implements the plane wave
pseudopotential method to solve the Kohn-Sham equations of 
DFT \cite{Hohenberg64,Kohn65,Perdew92,White94}
with respect to the Schr\"odinger Hamiltonian,
to calculate energies and Hellmann-Feynman forces and stresses.

Pseudopotentials were generated using the Troullier-Martins method
\cite{Troullier91} with 1s2s2p (Al) and 1s2s2p3s3p (Ni) electron
shells treated as core. Consequently 13 valence electrons per unit
cell are considered. NiAl was treated as non-magnetic.

For a cubic material, a sequence of constant volume calculations suffices
to determine the cold compression curve.  The volume was held fixed
while restoring forces were calculated to determine the dynamical
matrix \cite{Ackland97}.

At standard temperature and pressure (STP), 
NiAl adopts the CsCl (or B2) structure, with lattice parameter 
$2.8870\pm 0.0001$\,\AA\ \cite{Miracle93},
giving a crystal density of 5.912\,g/cm$^3$.
A plane wave cutoff of 900\,eV and a symmetry-reduced 10$^3$ $k$-point grid
\cite{Monkhorst76} was sufficient to converge the Pulay-corrected ground states
\cite{Pulay69,Francis90,Warren96,Hsueh96} to $\sim$10\,meV/\AA$^3$ or better.
This level of convergence in pressure was 1-2 orders of magnitude smaller than
the discrepancy from measured pressures introduced through the use of DFT.

\subsection{Isotropic compression}
\begin{figure}
\begin{center}\includegraphics[scale=0.72]{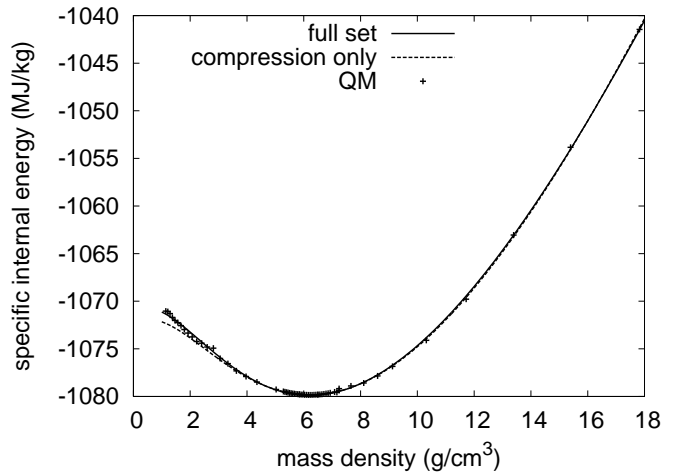}\end{center}

\caption{Quantum mechanical predictions of cold compression curve (energy),
   and fit with Rose function.}
\label{fig:eccmp}
\end{figure}

\begin{figure}
\begin{center}\includegraphics[scale=0.72]{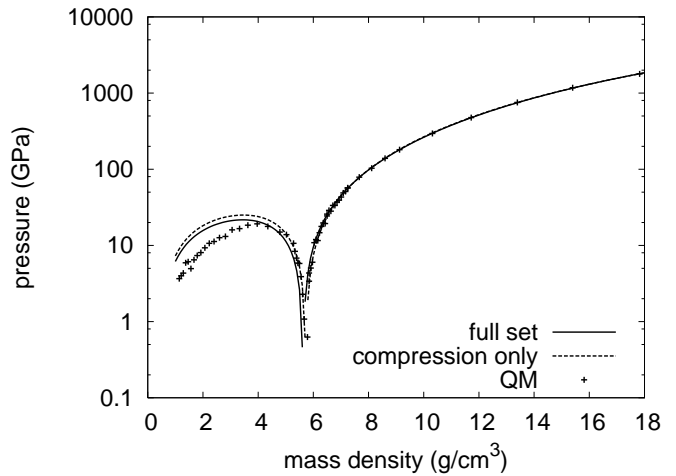}\end{center}

\caption{Quantum mechanical predictions of cold compression curve (pressure),
   and fit with Rose function (logarithmic scale).}
\label{fig:pclogcmp}
{\small The absolute value of pressure is plotted in regions of tension.}
\end{figure}

Isotropic compression calculations were performed to determine the
cold compression curve $e_c(\rho)$.
The lattice parameter was varied between 2.0 and 5.0\,\AA\
at intervals of 0.1\,\AA, with additional calculations at intervals of
0.05\,\AA\ between 2.5 and 3.1\,\AA\ and
0.01\,\AA\ between 2.7 and 3.0\,\AA.
(Figs~\ref{fig:eccmp} and \ref{fig:pclogcmp}.)

It is instructive to fit the electronic structure calculations to the
Rose functional form \cite{Rose84}, which has been found empirically
to describe the compression behavior of many elements.
A functional fit such as the Rose form is useful in that it provides
a much more compact representation of the cold curve than does the
tabulated relation obtained from a series of electronic structure
calculations.
It is interesting to assess the accuracy of this functional form in
reproducing the expansion region as well as compressed states.
For use with atomic structure calculations, in which the
binding energy contains a contribution from electrons to atoms as well as
atoms to form the solid, the variation of binding energy with
scaled compression can be described by
\begin{equation}
e=e_0-e_1(1+a+0.05a^3)e^{-a},
\end{equation}
which is a slight generalization of the original form of the relation
\cite{Sutton93}.
The parameter $a$ is a scale length defined
with respect to the Wigner-Seitz radius $r_{WS}$
as a function of specific volume,
\begin{equation} 
a(v) = \dfrac{r_{WS}(v)-r_{WS}(v_0)}l.
\end{equation}
The pressure along the cold compression curve is then
\begin{equation}
p=-3B_0\dfrac{(v/v_0)^{1/3}-1}{(v/v_0)^{2/3}}(1-0.15a+0.05a^2)e^{-a}.
\end{equation}
The zero-pressure bulk modulus is related to the binding energy scale by
\begin{equation}
B_0 = \dfrac{e_1 m_a}{12\pi r_{WS}(v_0) l^2}.
\end{equation}
The cold curve for a material is then described by three principal parameters
-- $\rho_0=1/v_0$, $e_1$ or $B_0$, and $l$ -- and an energy offset $e_0$.
The Rose function has a theoretical connection with simple metals, in which
the compression properties are dominated by the electron density.
Here, it was regarded as a
convenient functional form for interpolating and smoothing data.

Non-linear optimization was used to obtain parameters from the 
quantum-mechanical
predictions of the frozen-ion cold curve in pressure-volume or
energy-volume space.
Some optimization schemes were unstable when the equilibrium volume $v_0$ was
included as a free parameter.
In practice, $v_0$ was first estimated by inspection, and
the remaining parameters ($B_0$, $l$, and $e_0$) were calculated for
fixed $v_0$.
This process was repeated for a few different values of $v_0$ to
investigate the sensitivity, and then a robust but inefficient Monte-Carlo 
optimization scheme was used to locate the optimum value of $v_0$.

The Rose function was not able to reproduce the shape of the quantum mechanical
cold compression curve over its whole range.
Separate fits were produced using the whole set of quantum mechanical states,
and using only the states under compression, which is the more relevant 
part of the curve for the shock wave studies of interest.
The fit to the compression portion was good: generally within the
numerical scatter of the quantum mechanical states.
It is possible that the Rose function performs worse in expansion because it
does not represent adequately the localization of electrons on individual atoms 
as the rarefied metal ceases to conduct.
The predicted parameters for NiAl were bracketed by the
(observed) parameters for Ni and Al, but not so as to suggest 
any systematic trend (e.g. averaging) that could be used to predict
alloy properties;
this highlights the inaccuracy of using mixture models
as is commonly attempted to predict
the properties of alloys, particularly intermetallic compounds
\cite{Trunin98}.
(Table~\ref{tab:rose} and Figs~\ref{fig:eccmp} to \ref{fig:pclogcmp}.)

\begin{table*}
\caption{Rose parameters fitted to
   {\it ab initio} frozen-ion cold compression curve for NiAl.}
\label{tab:rose}
\begin{center}
\begin{tabular}{|l|l|l|r|r|}\hline
 & $\rho_0$ & $l$ & $B_0$ & $e_0$ \\
 & (g/cm$^3$) & (\AA) & (GPa) & (MJ/kg) \\ \hline
full set & 6.200 & 0.2860 & 177.8 & -1069 \\
compression only & 6.227 & 0.2966 & 189.0 & -1068 \\
\hline
compression only, adjusted by 7\,GPa & 6.007 & 0.2934 & 164.8 & -1070 \\
\hline
Ni & 8.90 & 0.270 & 186.0 & - \\
Al & 2.70 & 0.336 & 72.2 & - \\
\hline\end{tabular}
\end{center}

{\small
Parameters for elemental Ni and Al \cite{Sutton93,Kittel96} 
are shown for comparison.
}
\end{table*}

%\begin{figure}
%\begin{center}\includegraphics[scale=0.72]{eclogcmp.eps}\end{center}
%\caption{Quantum mechanical predictions of cold compression curve (energy),
%   and fit with Rose function (logarithmic scale).}
%\label{fig:eclogcmp}
%\end{figure}

%\begin{figure}
%\begin{center}\includegraphics[scale=0.72]{pccmp.eps}\end{center}
%\caption{Quantum mechanical predictions of cold compression curve (pressure),
%   and fit with Rose function.}
%\label{fig:pccmp}
%\end{figure}

\subsection{Thermal excitation of the electrons}
At elevated temperatures, excitations of the electronic states can
contribute to the free energy.
The electron-thermal contribution $e_e(\rho,T)$ was calculated from
the Kohn-Sham electron band structure by populating the calculated states
according to Fermi-Dirac statistics, using the procedure applied previously
to other elements including Si \cite{Swift_sieos_01}.
The band structure was calculated at a temperature of 0\,K in the
frozen-ion approximation: the electron states were not re-calculated
self-consistently with the state occupations at finite temperatures.
Over the range of states considered, which included temperatures up to
10000\,K, the electron excitation was rather small.
This can be seen by considering the electron-thermal energy,
found to make only a small contribution to the EOS in this regime,
so any error in the band energies from the limitations of DFT or
electron-phonon coupling would be a small correction
to this small contribution.

The electron wavefunctions were represented at a finite set of
positions in reciprocal space, $\{\vec k_i\}$, reduced by
symmetry operations so states at each $k$-point had
a weight $w_i$.
The energy levels were used directly in estimating $e_e(v,T)$,
rather than collecting them into a numerical distribution function $g(E)$.

Given the set of discrete levels for each compression,
the chemical potential $\mu$ was found as a function of temperature $T$
by constraining the total number of valence electrons $N$:
\begin{equation}
\sum_i \frac{w_i}{e^{(E_i-\mu(T))/k_BT}+1} = N,
\end{equation}
using an iterative inversion algorithm.
Once $\mu(T)$ had been determined in this way, 
the expectation value of the electronic energy was calculated:
\begin{eqnarray}
\langle E(T)\rangle = \sum_i \frac{E_i w_i}{e^{(E_i-\mu(T))/k_BT}+1}.
\end{eqnarray}
Repeating this calculation for each compression, and dividing by the mass of the
atom to obtain the specific energy of excitation, the electron-thermal
contribution $e_e(\rho,T)$ to the EOS was obtained.

The electron-thermal energy was at least an order of magnitude smaller than the
lattice-thermal energy over the range of temperatures and compressions
considered.
At a fixed temperature, the energy decreased with compression as the bands 
broaden with respect to energy, decreasing $g(\mu)$.
The electron-thermal energy did however increase more rapidly with
temperature than did the lattice-thermal energy, so at temperatures above a few 
electron-volts the electron-thermal energy dominates.
(Fig.~\ref{fig:tecmp}.)

\subsection{Phonon modes}
The thermal motion of the atoms and its contribution to the EOS
were calculated in terms of the phonon modes of the lattice,
following the method described previously for Si \cite{Swift_sieos_01}.
In general, the restoring force for displacement each atom from
equilibrium may be anharmonic.
There is a large increase in computational complexity in calculating 
derivatives in the lattice potential energy beyond those required for
harmonic phonons.
In the present work, we have calculated effective quasiharmonic phonon
modes for relevant amplitudes of atomic displacement, and assessed the
sensitivity of the EOS to different choices of displacement.

Quasiharmonic phonons were calculated using a force-constant method
\cite{Ackland97,Ashcroft76,Maradudin63}.
{\it Ab initio} elements in the
dynamical matrix were obtained by calculating the force on all the
atoms when one atom was perturbed from its equilibrium position,
from the charge distribution in the quantum mechanical description of
the electron ground state.

When atom $i$ is perturbed by some finite
displacement $\vec u_i$ from its equilibrium position $\vec r_{0i}$,
the force on each atom can be found from the electron ground state.
If $\Phi$ is the electron ground state energy,
the calculation provides elements of the stiffness matrix ${\bf D}$,
between the displaced atom $i$ and all other atoms $j$ in the supercell.
\begin{eqnarray}
{\bf D}(\vec r_i - \vec r_j)\equiv
\pdiffl{^2\Phi}{\vec u_i\partial\vec u_j}\simeq
\pdiffl{\Phi(\alpha\hat u_i)}{\vec u_j}
\frac 1\alpha.
\end{eqnarray}
$\partial\Phi(\alpha\hat u_i)/\partial\vec u_j$ is the
force $\vec f_j$ on atom $j$ when atom $i$ is displaced
by a distance $\alpha$ in the direction (unit vector) $\hat u_i$.
Because partial differentiation operators commute for a smooth function,
a row and column of the eigenproblem can be 
determined from an electron ground state calculation with a displacement
$\vec u_i$ along one of the coordinate directions, by dividing the force
on each atom by the displacement.
Other elements can be generated using symmetry: for the CsCl 
it was sufficient to perturb the Ni atom at $(0,0,0)$ and the Al atom
at $(\frac 12,\frac 12,\frac 12)$ in turn along the $[100]$
direction in order to obtain the entire matrix of force constants.

The squared phonon frequencies are then the eigenvalues of
\begin{eqnarray}
\omega^2\vec u_i=\sum_j\pdiffl{^2\Phi}{\vec u_i\partial\vec u_j}.
e^{i\vec k.(\vec r_i-\vec r_j)}\frac{\vec u_i}{\sqrt{m_im_j}},
\end{eqnarray}
where $m_i$ is the mass of atom $i$.
The matrix on the right-hand side is the dynamical matrix, ${\bf \tilde D}$:
\begin{eqnarray} 
[{\bf\tilde D(\vec k)}]_{\alpha\beta}\equiv
\sum_{ij}\pdiffl{^2\Phi}{[\vec u_i]_\alpha\partial[\vec u_j]_\beta}.
e^{i\vec k.(\vec r_i-\vec r_j)}\frac 1{\sqrt{m_im_j}},
\end{eqnarray}
where square brackets are used to denote an element of a matrix.

Restoring forces were calculated and the dynamical matrix
constructed for several magnitudes of the displacement of the atoms.
This allowed the deviation from a harmonic potential to be estimated,
and can be used to determine some of the anharmonic components in the
potential.
The displacements chosen were $0$, $\pm 0.001$, and $0.01$ of the 
lattice parameter of the $2\times 2\times 2$ supercell used.
The restoring forces were calculated for $a=3.1$, $2.9$, $2.7$, $2.5$, $2.3$,
and $2.1$\,\AA \ for the CsCl primitive cell.
The potential was found to be linear in the magnitude of displacement,
to around $\pm 0.005$\,eV/\AA.
This linearity indicates that the potential surface experienced by the
atoms is harmonic, an assertion which was investigated by constructing EOS
based on different magnitudes of displacement.
% (Fig.~\ref{fig:fcmp}).

%\begin{figure}
%\begin{center}\includegraphics[scale=0.72]{fcmp.eps}\end{center}
%\caption{Linearity of restoring force on Ni atom at $(0,0,0)$
%   calculated with different displacements for a CsCl lattice parameter of
%   2.9\,\AA.}
%\label{fig:fcmp}
%\end{figure}

To calculate the density of phonon states $g(\omega)$
for a structure with a given set of lattice parameters,
the phonon eigenproblem -- diagonalizing
${\bf\tilde D}(\vec k)$ -- was solved for each of a set of wavevectors $\vec k$.
These were chosen randomly with a uniform distribution over
the Brillouin zone, by choosing the components to be three independent
random numbers with uniform distribution between 0 and 1.
As was found previously \cite{Swift_sieos_01}, 
the density of states converged slowly with the number of wavevectors.
We did not attempt to calculate fully-converged densities of states,
as previous experience showed that, as an integrated property of the
density of states, the EOS was only sensitive to low moments of the
distribution.
These moments did not change significantly with compression.
%(Figs~\ref{fig:dos0.001} and \ref{fig:dos0.01}.)

%\begin{figure}
%\begin{center}\includegraphics[scale=0.72]{dos0.001.eps}\end{center}
%\caption{Density of phonon states predicted at different values of the
%   lattice parameter, for a displacement of 0.001 times the supercell
%   lattice parameter.}
%\label{fig:dos0.001}
%\end{figure}

%\begin{figure}
%\begin{center}\includegraphics[scale=0.72]{dos0.01.eps}\end{center}
%\caption{Density of phonon states predicted at different values of the
%   lattice parameter, for a displacement of 0.01 times the supercell
%   lattice parameter.}
%\label{fig:dos0.01}
%\end{figure}

\begin{figure}
\begin{center}\includegraphics[scale=0.72]{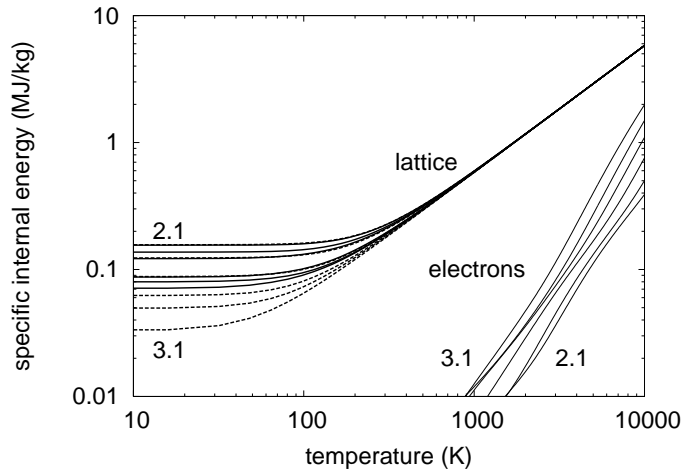}\end{center}

\caption{Thermal energies.
   The lattice-thermal energy was calculated from phonon densities of states
   which were deduced using different atomic displacements
   (solid: displacement of 0.001; dashed: 0.01):
   3.1, 2.9, 2.7, 2.5, 2.3, and 2.1\,\AA\ moving up the vertical axis
   as the zero-point energy increases with compression.
   The electron-thermal energies were calculated by populating the
   zero-temperature band structure.
   Note that the electron-thermal energy decreases with compression,
   as the bands move to higher energies.}
\label{fig:tecmp}
Note the finite and compression-dependent value of the Debye temperature,
at which all curves asymptote to the same straight line.
\end{figure}

The variation of lattice-thermal energy with temperature was found
by populating the phonon modes according to Boltzmann
statistics \cite{Ashcroft76}.
At a given mass density, the lattice thermal energy
is
\begin{eqnarray}
E_l(T)=
\sum_i g(\omega_i)\hbar\omega_i
\left(\dfrac 1{e^{\hbar\omega_i/kT}-1} + \dfrac 12\right),
\end{eqnarray}
from which the lattice-thermal contribution $e_l(v,T)$ to the EOS was found 
by normalizing to 3 modes per atom.
(Fig.~\ref{fig:tecmp}.)

Although the phonon modes could be calculated reliably around zero pressure
as has been found previously \cite{Huang04},
some negative eigenvalues of the dynamical matrix were found
at higher compressions, giving a small proportion of 
imaginary phonon frequencies.
Imaginary frequencies indicate possible instabilities in the
lattice: the restoring force on a displaced atom does not increase
with displacement, so the displacement can grow.  This situation may
be related to a major phase transition,
a minor perturbation of the structure 
which is dynamically stabilized at finite temperature \cite{Pinsook99}, 
a restoring force that is non-linear in displacement,
or it may be caused by minor numerical inconsistencies when some 
force components which should be identical by symmetry are calculated 
redundantly in different directions or by displacing different atoms.
Infinitesimal displacements will not
distinguish between these possibilities, though finite displacements
may be used to investigate the structure of the potential field
experienced by each atom.  In either case, the local potential is no
longer quadratic, and the quasiharmonic model is not strictly valid.
Minor perturbations of the structure and non-linear restoring forces
can be treated formally by renormalization of the phonon modes 
\cite{Drummond01}, though this procedure is unwieldy to apply 
consistently with the
detailed shape of the restoring force at finite displacements.
In constructing the EOS for Si \cite{Swift_sieos_01}, where nonlinear
contributions to the restoring force were more pronounced,
the principal shock Hugoniot was found to be insensitive to details of
the phonon density of states, including the treatment of imaginary frequencies.
We investigated the sensitivity of the NiAl EOS to different methods for
taking account of the imaginary frequencies:
treating them as freely translational -- contributing a heat
capacity of $k_B/2$ at all temperatures -- or ignoring them and 
renormalizing the density of states to 3 modes per atom.
As shown below,
the difference in shock properties was negligible for pressures up to 70\,GPa.

\subsection{Construction of the equation of state}
For computational convenience, the zero-point energy of the lattice modes
was subsumed in the lattice-thermal contribution rather than the cold curve;
thus the cold curve strictly assumed frozen ions.
This assumption does not affect the accuracy of the resulting EOS, but
the cold curve is not physically real as defined.
The electron-thermal energy can be predicted from band structure
calculations closely related to the electron ground state calculations.
However, we have found for many materials that the electron-thermal energy
makes a negligible contribution to the EOS of the solid,
and the contribution in NiAl has been calculated to be relatively unimportant
\cite{Wang04}, so it was ignored in the present work.

\subsubsection{Thermodynamic completion}
Given the total specific energy $e(T)$ along each isochore,
the specific entropy $s$ was found by
integration of the second law of thermodynamics ($de=Tds-pdv$):
\begin{eqnarray} 
s(T) = \int_0^T\dfrac{dT'}{T'}\pdiffl e{T'}.
\end{eqnarray}
The (specific) free energy $f$ was then calculated from $f=e-Ts$,
and the pressure $p$ was calculated by differentiating $f$.
The thermodynamic functions were represented by tables,
so local polynomials were fitted through adjacent sets of points
-- generally quadratics through sets of three points -- to
allow differentiation and integration to be performed.

EOS were generated in SESAME 301 format \cite{SESAME}.
This consists of rectangular tables of pressure $p$ (GPa) and specific internal
energy $e$ (MJ/kg) as functions of temperature $T$ (K) and density $\rho$
(\SIdens).
For each of the densities in the original cold curve calculations, states were
calculated along an isochore from $T=0$ to 10000\,K.
Although the most straightforward form to generate, SESAME table 301 is rather
inconvenient for hydrodynamic calculations. A standard hydrocode requirement is
for an EOS of the form $p(\rho,e)$. To derive this from a table 301 requires
$T$ to be found by inverse interpolation given $\rho$ and $e$, and then $T$ and
$\rho$ used to find $p$.
In the present work, bilinear interpolation was used in finding $p$ and $e$ at
states between the ordinates of $\rho$ and $T$.

\subsubsection{Adjustment to reproduce STP mass density}
As described previously \cite{Swift_sieos_01}, 
any EOS can be corrected to reproduce the observed STP
state by adding a pressure offset $\Delta p_c$ and for consistency a
corresponding energy tilt $\Delta e_c = -v\Delta p_c$.
We have found that the pressure offset improves the agreement with
compression data better than do other types of adjustment, such as 
scaling of the mass density.
However, the quantum-mechanical cold curves asymptoted correctly toward
zero pressure as the lattice parameter became large, so a constant 
pressure offset would not be accurate at low density.
The pressure offset was calculated from the pressure given by each EOS
at 293\,K and the observed STP crystal density of 5.912\,g/cm$^3$
(Table~\ref{tab:poffsetandbulkmod}).
A modified Rose fit was calculated for the corrected cold curve
(Table~\ref{tab:rose}).
Of course, this fit does not reproduce the STP mass density because
thermal expansion is not included.
One straightforward test of an EOS is to calculate the bulk modulus.
For NiAl, the bulk modulus has been measured to be $156\pm 3$\,GPa
\cite{Otto97} and as inferred from the elastic constants is 166.0\,GPa
\cite{Wasilewski66}.
The adjusted first principles -- {\it ab fere initio} -- EOS 
gave bulk moduli which lay close to these values
(Table~\ref{tab:poffsetandbulkmod}).

\begin{table}
\caption{Pressure offset and bulk modulus
   for {\it ab fere initio} equations of state.}
\label{tab:poffsetandbulkmod}
\begin{center}
\begin{tabular}{|l|r|r|}\hline
{\bf atom} & {\bf offset} & {\bf bulk modulus} \\
{\bf displacement} & {\bf (GPa)} & {\bf (GPa)} \\ \hline
0.001 & -7.21 & 165.6 \\
0.01 & -6.65 & 154.0 \\
\hline\end{tabular}
\end{center}
\end{table}

\section{Isothermal compression}
\begin{figure}
\begin{center}\includegraphics[scale=0.72]{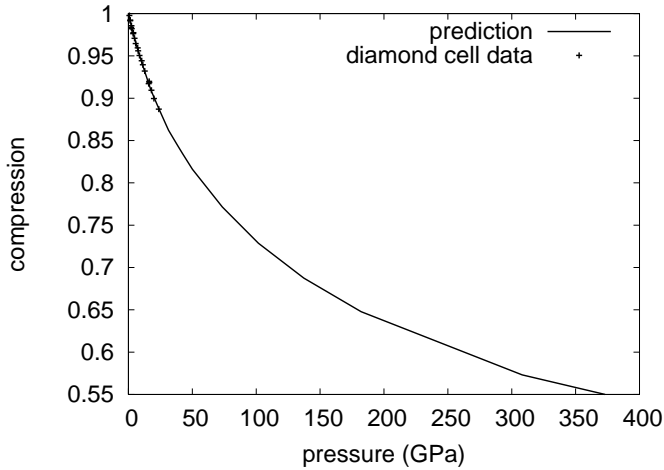}\end{center}

\caption{Comparison between calculated ambient isotherms and diamond anvil data.
   Alternative theoretical EOS gave isotherms which were not significantly
   different.}
\label{fig:ambiso}
\end{figure}

\begin{figure}
\begin{center}\includegraphics[scale=0.72]{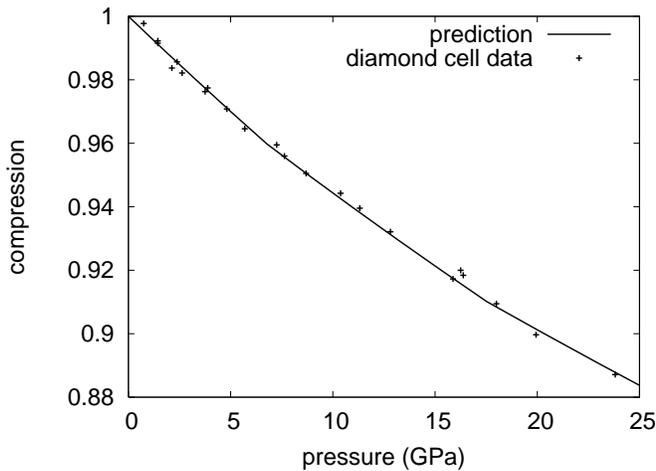}\end{center}

\caption{Comparison between calculated ambient isotherms and diamond anvil data
   (detail at lower pressures).}
\label{fig:ambiso1}
\end{figure}

Diamond-anvil measurements have been reported for the isothermal
compression of NiAl to 25\,GPa \cite{Otto97}.
Isothermal compression predictions were extracted directly from
the $p(\rho,T)$ table.
Isotherms from the EOS based on the ground state energies passed through
the diamond-anvil data within its scatter.
(Figs~\ref{fig:ambiso} and \ref{fig:ambiso1}.)

\section{Shock compression}
The theoretical EOS were used to predict the principal shock Hugoniot,
by solving the Rankine-Hugoniot equations \cite{Bushman93}
linking the states on either side of the shock to its velocity $u_s$:
\begin{eqnarray}
u_s^2 & = & v_0^2\dfrac{p-p_0}{v_0-v} \\
u_p & = & \sqrt{\left[(p-p_0)(v_0-v)\right]} \\
e & = & e_0 + \frac 12 (p+p_0)(v_0-v)
\end{eqnarray}
where subscript `0' denotes material ahead of the shock (with $u_p=0$).
Given the EOS in the form $p(v,e)$ this set of equations can be closed, 
allowing the
Hugoniot (locus of states reached from the initial state by a single shock)
to be calculated.
The phonon and electron modes were populated at each mass density and 
temperature in the construction of the EOS, so the Rankine-Hugoniot equations
were in effect solved self-consistently with the population of the thermal 
modes, though this was done indirectly through the use of the EOS.

Like the EOS itself, the shock Hugoniot was calculated up to pressures of
several hundred gigapascals.
The comparison with isothermal compression data demonstrates the accuracy
of the EOS, and the electron band-structure remains valid until
the pseudopotentials start to overlap, which typically requires pressures
in the terapascal regime.
As well as providing an {\it a priori} prediction of the EOS and Hugoniot,
the high-pressure calculations serve to explore the sensitivity of the 
theoretical predictions to different assumptions and simplifications used
in constructing EOS, and therefore where greater care -- or experiment -- 
is needed to constrain the EOS.

\begin{figure}
\begin{center}\includegraphics[scale=0.72]{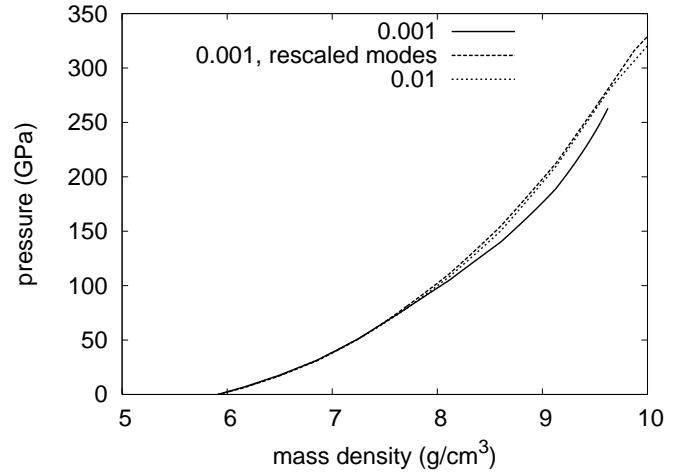}\end{center}

\caption{Comparison between principal shock Hugoniot from different
   phonon treatments, in density -- pressure space.
   The atom displacement is expressed with respect to the
   superlattice parameter.}
\label{fig:hugdp1}
\end{figure}

\begin{figure}
\begin{center}\includegraphics[scale=0.72]{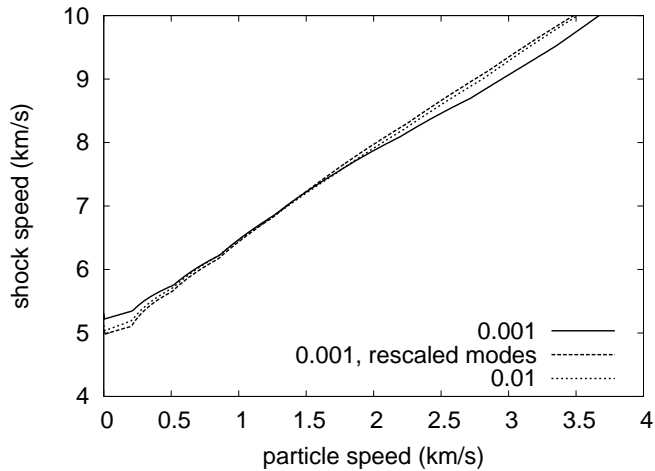}\end{center}

\caption{Comparison between principal shock Hugoniot from different
   phonon treatments, in particle speed -- shock speed space.
   The atom displacement is expressed with respect to the
   superlattice parameter.}
\label{fig:hugupus1}
\end{figure}

\begin{figure}
\begin{center}\includegraphics[scale=0.72]{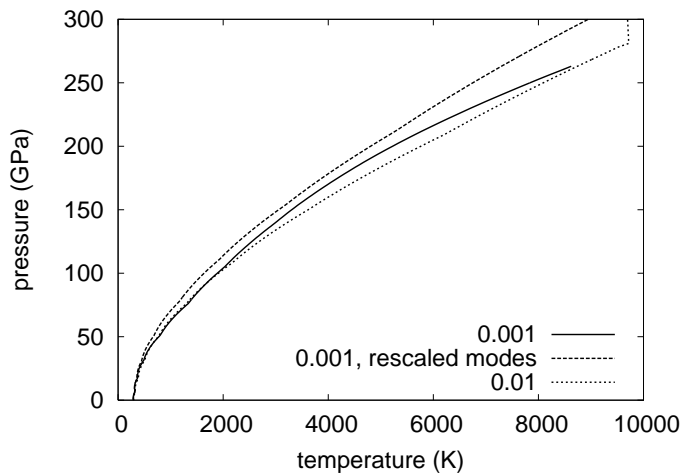}\end{center}

\caption{Comparison between principal shock Hugoniot from different
   phonon treatments, in temperature -- pressure space.
   The atom displacement is expressed with respect to the
   superlattice parameter.}
\label{fig:hugtp1}
\end{figure}

The quasiharmonic EOS showed some sensitivity to the atom displacement and the 
treatment of the imaginary modes.
The variation is an indication of the uncertainty in the theoretical EOS,
and of regimes in which anharmonic terms in the lattice-thermal energy 
(phonon-phonon interactions) may contribute significantly.
In pressure-density space, the Hugoniots were very similar below around 80\,GPa,
then differed by around 15\%\ at higher pressures.
In shock speed-particle speed space, the Hugoniots varied by 5\%\ at low
pressures, became coincident for shock speeds around 7\,km/s
(60\,GPa), then deviated by around 5\%\ at higher pressures.
The deviation was most pronounced in pressure-temperature space:
around 15\%.
With the density of phonon states corrected to remove imaginary modes
-- labeled as `rescaled' in the figures -- the principal Hugoniot for
a displacement of 0.001 lay much closer to that for 0.01.
(Figs~\ref{fig:hugdp1} to \ref{fig:hugtp1}.)

\begin{figure}
\begin{center}\includegraphics[scale=0.72]{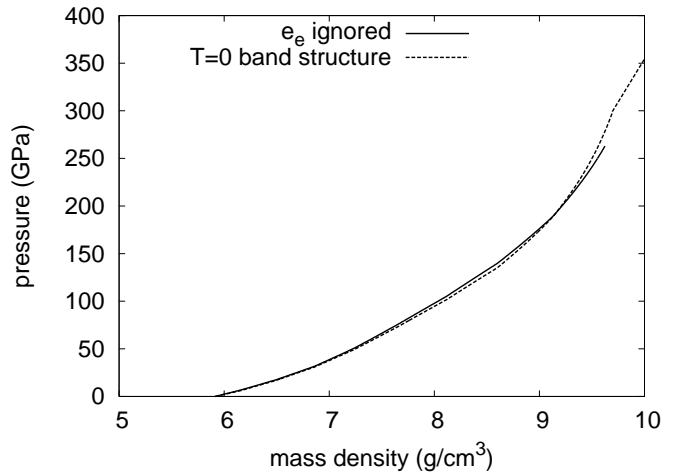}\end{center}

\caption{Comparison between principal shock Hugoniot from different
   electron-thermal treatments, in density -- pressure space.
   The phonons were calculated with a displacement of 0.001 times the
   superlattice parameter, without rescaling.}
\label{fig:hugetdp1}
\end{figure}

\begin{figure}
\begin{center}\includegraphics[scale=0.72]{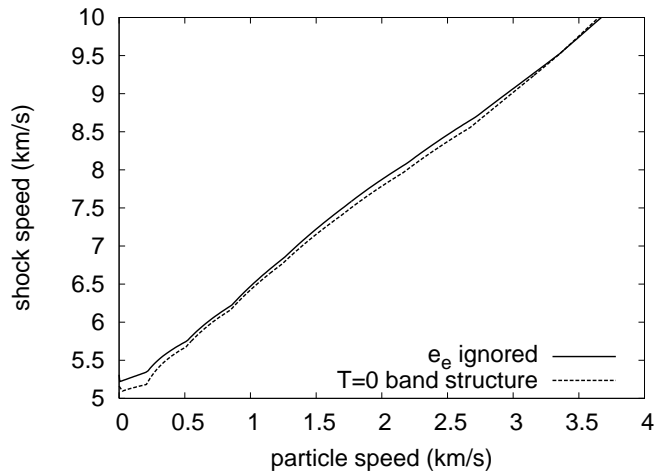}\end{center}

\caption{Comparison between principal shock Hugoniot from different
   electron-thermal treatments, in particle speed -- shock speed space.
   The phonons were calculated with a displacement of 0.001 times the
   superlattice parameter, without rescaling.}
\label{fig:hugetupus1}
\end{figure}

\begin{figure}
\begin{center}\includegraphics[scale=0.72]{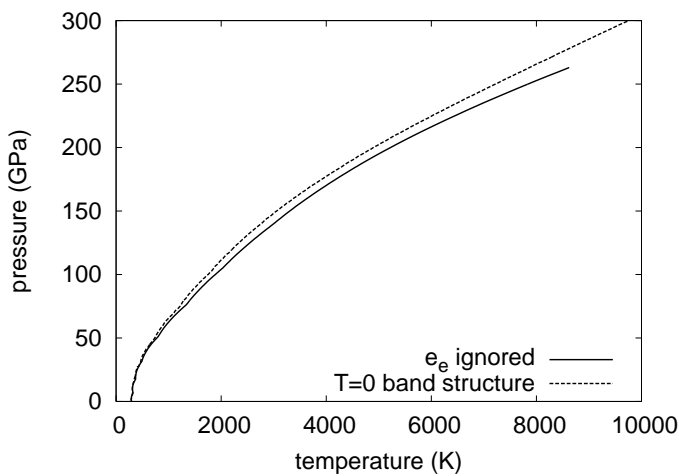}\end{center}

\caption{Comparison between principal shock Hugoniot from different
   electron-thermal treatments, in temperature -- pressure space.
   The phonons were calculated with a displacement of 0.001 times the
   superlattice parameter, without rescaling.}
\label{fig:hugettp1}
\end{figure}

The EOS showed little sensitivity to the inclusion of the electron-thermal
contribution, except in pressure-temperature space where it rose linearly
with temperature, reaching around 10\%\ in pressure by 10000\,K.
The exaggerated effect in temperature is not surprising: over the
regime studied the EOS is dominated by the cold compression curve,
so a given, relatively small, difference in thermal pressure equates
to a considerably larger difference in temperature.
(Figs~\ref{fig:hugetdp1} to \ref{fig:hugettp1}.)

\section{Shock wave experiments}
The experiments described here were intended to help validate the
theoretical EOS and to calibrate the model of flow stress.
In many cases the samples were recovered after the experiments,
so metallographic analysis could be performed in future if desired.

To generate shock states by impact, 
the TRIDENT laser was used to launch flyer plates, 
each of which then impacted a stationary target.
Two series of experiments were performed.
In the first, the flyer was Cu and the target comprised a crystal of NiAl,
sometimes releasing into a polymethyl methacrylate (PMMA) window.
In the second series, the flyer was NiAl and the target a transparent LiF
crystal.
The flyer speed and surface velocity of the sample were measured
by laser Doppler velocimetry.

The laser pulse was 600\,ns long, and the flyers were 
around 50 to 400\,$\mu$m thick.
Velocities were a few hundred meters per second with a laser energy
of 5 to 20\,J over a spot 5\,mm in diameter.

The accuracy of the laser flyer technique was previously evaluated in 
experiments on the EOS of Cu \cite{Swift_flyeracc_05},
and has also been used to measure the EOS of NiTi \cite{Swift_NiTiEOS_05}.
The experiments reported here were performed at TRIDENT
as part of the `Pink Flamingo' (December 2001) and `Flying Pig' (March 2002)
campaigns.

\subsection{Sample preparation}
Single crystal samples of NiAl were grown from the melt using the optical
floating zone technique.  Feed-rods were prepared via arc melting, with excess
Al included in the initial charge to compensate for losses during rod
preparation and crystal growth.  Crystals were grown at 15\,mm/hr along
$\langle 001\rangle$, starting from an oriented seed.
To compensate for the greater evaporation rate of Al, 
the initial melt was prepared with a slight excess of Al.
It is difficult to predict the evaporation rate precisely, and the
crystals were slightly Al-rich.
Single crystal NiAl was also obtained from
the General Electric Corp; this material was closer to stoichiometry.
The orientation was determined by back-reflection Laue diffraction.
Samples were sliced parallel to $(100)$ and $(110)$ planes, 
then ground and polished 
to the desired thickness using diamond media, to a 1\,$\mu$m mirror finish. 
The grinding process imparted some pre-strain close to the polished surfaces.

\subsection{Experimental configurations}
\begin{figure}
\begin{center}
\includegraphics[scale=0.5]{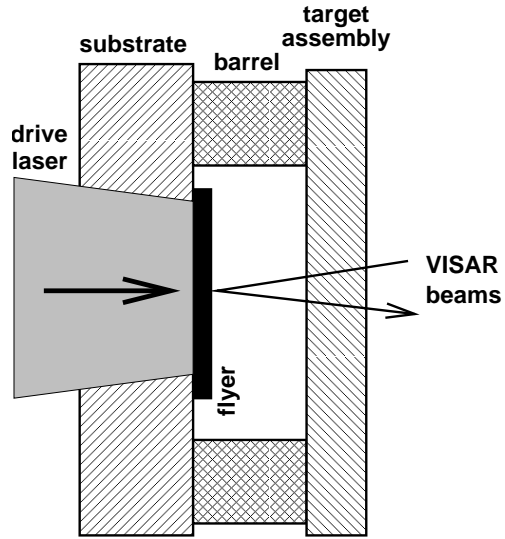}
\end{center}

\caption{Schematic cross-section of laser-launched flyer impact experiments.}
\label{fig:flyerschem}
\end{figure}

Several different configurations of flyer experiment were used.
In all cases, the flyer was attached to its transparent substrate and spaced
off from the target assembly by a `barrel' comprising a stack of plastic shims.
The barrel was typically around 500\,$\mu$m long, allowing enough space for the 
flyer to accelerate before impact.
(Fig.~\ref{fig:flyerschem}).

\begin{figure}
\begin{center}
\includegraphics[scale=0.4]{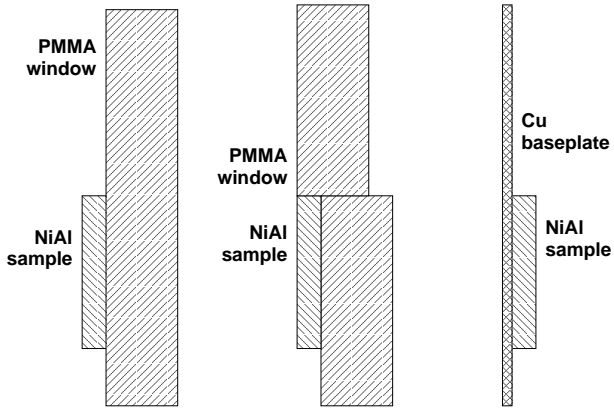}
\end{center}

\caption{Schematic of target assembly for PMMA release
   and baseplate configurations.
   Center: stepped window providing better measurement of impact time.
   In all cases the flyer approaches from the left.}
\label{fig:exptschem}
\end{figure}

In the initial experiments with a Cu flyer and NiAl target, 
the sample covered only half of the flyer,
so the flyer could be seen and its velocity measured.
Several variants were used in the design of the target assembly,
to investigate the accuracy of data from each.
In most of these experiments, the NiAl sample was in contact with a PMMA
release window; in some of the
PMMA window experiments, the window was stepped to provide a more accurate
measurement of the time at which impact occurred and thus the shock transit
time through the sample. 
Experiments were also performed in which the sample was mounted on a Cu
baseplate: the baseplate obscured the view of the flyer, but shock breakout
at the surface of the baseplate provided a measurement of the pressure
(from the peak free surface particle speed) and gave a relatively accurate
measurement of the time at which the shock entered the sample.
The baseplate design also avoided difficulties from light reflected from the
free surface of the PMMA windows, which sometimes obscured the signal from
the flyer or the sample.
The baseplate had to be relatively thin to avoid decay of the shock,
and the sample was not recovered from experiments using this design.
(Fig.~\ref{fig:exptschem}.)

The flyer speeds were low compared with the sound speed in the target
materials, so the finite accuracy of assembly -- several micrometers in
co-planarity or thickness -- and the finite temporal registration between the
point and line VISARs resulted in large uncertainties in shock transit times.
The second series of experiments was designed to minimize the uncertainty
in inferred particle speed and shock pressure: by using the deceleration of
a NiAl flyer with a window of known properties (LiF), the Hugoniot state
could be inferred from the point VISAR record only, the less precise 
line VISAR being used as a coarse velocity measurement (e.g. to help count
fringe jumps in the point VISAR record) and to verify flatness.

\subsection{Target assembly}
PMMA substrates were used, coated with layers of
C, Al, and Al$_2$O$_3$, in order to absorb the laser energy
and insulate the flyer from heating \cite{Swift_flyeracc_05}.
Cu flyers were punched from foils purchased from Goodfellow Corp.
The foils had striations and machining marks which could generate
interference patterns that could confuse the interpretation of the
laser velocimetry records.
The foils were polished manually using diamond paste to reduce the regularity
of the marks.
The material for the NiAl flyers was roughly triangular in shape.
Rather than cutting or punching flyer disks before assembly, the triangles
were attached whole to the substrates, and the drive laser punched out the
central 5\,mm to form the flyer.
The material remaining attached to the substrate formed a seal which prevented
plasma from the drive escaping radially.
The barrel was much shorter than the flyer diameter, so the attachment of the
edges did not make the central portion of the flyer curve measurably
and did not reduce its speed.

\begin{figure}
\begin{center}
\includegraphics[scale=0.65]{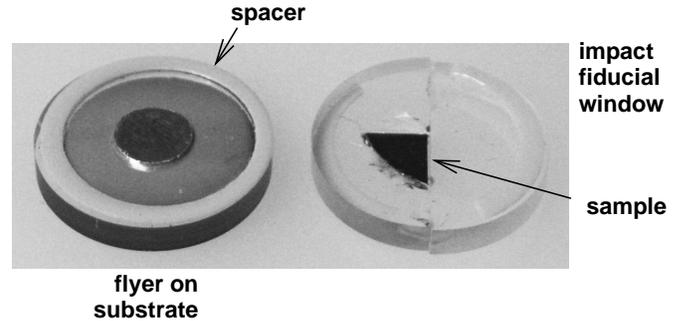}
\end{center}

\caption{Example of components of shock experiment during assembly
   (stepped release window design).  The flyer is 5\,mm in diameter;
   the sample is visible {\it through} the window.}
\label{fig:stepdisc_exploded}
\end{figure}

\begin{figure}
\begin{center}
\includegraphics[scale=0.60]{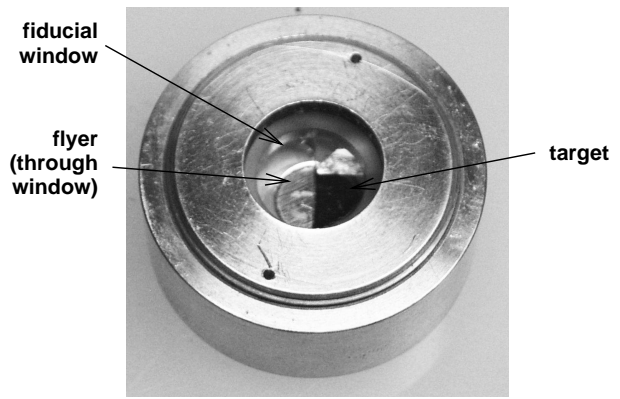}
\end{center}

\caption{Example of assembled components (stepped release window design).}
\label{fig:stepdisc_assembled}
\end{figure}

Components were glued together with five-minute epoxy.
Each flyer was clamped to its substrate while the glue set,
to minimize the thickness of the glue layer.
The initial viscosity of the glue was low, so the glue thickness was
estimated to be negligible, essentially filling in surface irregularities.
Components of the target assembly were clamped, and glued together by small
drops at their corners or edges to avoid introducing any layer of glue
between components which would change the shock and optical properties.
(Figs~\ref{fig:stepdisc_exploded} and \ref{fig:stepdisc_assembled}.)

\subsection{Diagnostics}
Laser Doppler velocimetry of the `velocity interferometry for surfaces of
any reflectivity' (VISAR) type \cite{Barker72} was used to measure the
velocity histories of the flyer and the sample.
A point-VISAR and a line-imaging VISAR were used simultaneously,
the point-VISAR signal being recorded on digitizing oscilloscopes and
the line-VISAR signal on an optical streak camera.
The point VISAR operated at a wavelength of 532\,nm with a 
continuous-wave source,
and the line VISAR at a wavelength of 660\,nm with a pulsed source
$\sim$1.5\,$\mu$s long.
Timing markers were incorporated on the streak record, at intervals
of 200\,ns, to provide a temporal fiducial and to allow non-linearities
in camera sweep to be removed.
Relative timing of the point and line VISARs was deduced by comparing
the position at which a shock wave appeared in a flyer impact experiment;
the relative timing had an uncertainty of $\sim$3\,ns.
The point VISAR was generally used to observed the NiAl.

\subsection{Drive beam}
The TRIDENT laser was operated in long-pulse mode, as in the previous flyer work
\cite{Swift_flyeracc_05,Swift_NiTiEOS_05}, using an acousto-optical
modulator to reduce the rate at which the pulse intensity rose and to
control its shape.
The drive pulse was chosen to be $\sim$600\,ns long (full width, half maximum),
and was delivered at the fundamental wavelength of the laser: 1054\,nm
(infra-red).
The pulses generated were asymmetric in time, with a long tail.

An infra-red random-phase plate (RPP) was added to smooth the beam;
this made a significant difference to the spatial uniformity.
The beam optics were arranged to give a spot 5\,mm in diameter
on the substrate.
The drive energy was quite low, so no RPP shield was included.
The RPP collected a small amount of debris from the substrate,
but was not significantly damaged.

\subsection{Results}
Eight experiments were performed with Cu flyers impacting NiAl,
and seven with NiAl flyers impacting LiF windows (Table~\ref{tab:shots}).
On one shot (14138), poor fringe contrast meant that the reference velocity 
could not be measured by velocimetry, so it was deduced from the 
laser energy instead.

\begin{table*}
\caption{Flyer impact experiments.}
\label{tab:shots}
\begin{center}\begin{tabular}{|c|c|r|rrr|r|rrr|}\hline
{\bf Shot} &
{\bf NiAl} &
   \multicolumn{4}{|c|}{{\bf Driver}} &
   \multicolumn{4}{|c|}{{\bf Target}} 
   \\ \cline{3-10}
 & {\bf orientation} & {\bf thickness} & \multicolumn{3}{|c|}{\bf speed} 
 & {\bf thickness} & \multicolumn{3}{|c|}{\bf speed} \\
 & & ($\mu$m) & \multicolumn{3}{|c|}{(m/s)}
   & ($\mu$m) & \multicolumn{3}{|c|}{(m/s)} \\ \hline
{\bf release window} &&&&&&&&& \\
14126 & (100) & 105 & 329 & $\pm$ & 15 & 200 & 331 & $\pm$ &  3 \\
14127 & (100) & 105 & 432 & $\pm$ & 20 & 217 & 390 & $\pm$ &  5 \\
14128 & (100) & 105 & 225 & $\pm$ & 10 & 217 & 232 & $\pm$ &  2 \\
\hline
{\bf baseplate} &&&&&&&&& \\
14136 & (100) & 250 & 165 & $\pm$ & 10 & 398 & 208 & $\pm$ & 2 \\
14137 & (100) & 250 & 200 & $\pm$ & 10 & 398 & 260 & $\pm$ & 2 \\
14138 & (100) & 250 & 185 & $\pm$ & 15 & 398 & 195 & $\pm$ & 5 \\
\hline
{\bf stepped disk} &&&&&&&&& \\
14139 & (100) & 250 & 153 & $\pm$ &  2 & 398 & 120 & $\pm$ & 10 \\
14140 & (100) & 250 & 180 & $\pm$ &  2 & 398 & 175 & $\pm$ & 10 \\
\hline
{\bf window impact} &&&&&&&&& \\
14381 & (100) & 94 & 325 & $\pm$ & 3 & - & 226 & $\pm$ & 2 \\
14389 & (100) & 375 & 102 & $\pm$ & 1 & - &  72 & $\pm$ & 1 \\
14390 & (100) & 389 & 152 & $\pm$ & 3 & - & 107 & $\pm$ & 3 \\
14406 & (100) & 350 & 141 & $\pm$ & 1 & - & 101 & $\pm$ & 1 \\
14387 & (110) & 228 & 167 & $\pm$ & 2 & - & 117 & $\pm$ & 2 \\
14388 & (110) & 262 & 167 & $\pm$ & 2 & - & 116 & $\pm$ & 2 \\
14405 & (110) & 392 & 149 & $\pm$ & 2 & - & 106 & $\pm$ & 1 \\
\hline\end{tabular}\end{center}

{\small
The baseplates were 55\,$\mu$m copper.
The driver speed is the peak free surface speed of the baseplate,
and the flyer speed on impact otherwise.
The impact window was a LiF crystal, 2\,mm thick,
with $(100)$ planes parallel to the impact surface.
The `target speed' in the window impact experiments is the speed of the
interface between the NiAl flyer and the LiF window immediately after impact.
}
\end{table*}

The drive energy was measured with a calorimeter,
and the irradiance history of the drive pulse with a photodiode.
The uncertainty in energy was of the order of 1\,J.
The pulse shape was repeatable at the same and different energies.

The VISAR records were used to measure the velocity history
and flatness of each flyer.
The flyers were flat to within the accuracy of the data over the central 3\,mm,
with a slight lag at the edges.
In some experiments, we recorded the impact of the flyer with a window
and thus were able to measure the flatness directly after several hundred
microns of flight.
Most of the flyers were still accelerating slightly at the end of the record.
There was evidence of ringing during acceleration, but no sign of
shock formation or spall in the flyers.

\begin{figure}
\begin{center}
\includegraphics[scale=0.72]{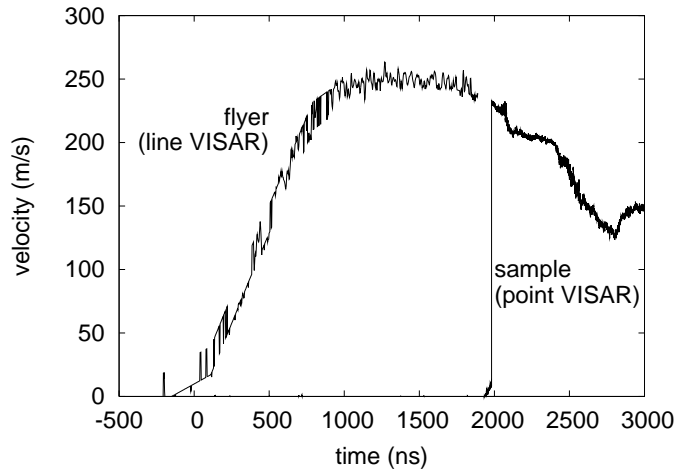}
\end{center}

\caption{Example velocity history from a window release experiment
   (shot 14128).}
\label{fig:14128_vcmp}
\end{figure}

\begin{figure}
\begin{center}
\includegraphics[scale=0.72]{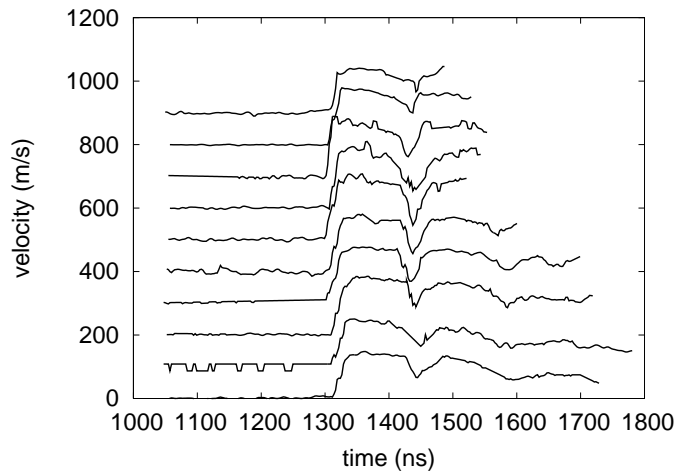}
\end{center}

\caption{Example spatially-resolved velocity history from a window release
   experiment (shot 14140: stepped release window).}
\label{fig:14140_vline}
{\small
The velocity from successive fringes has been displaced by 100\,m/s for clarity.
}
\end{figure}

\begin{figure}
\begin{center}
\includegraphics[scale=0.72]{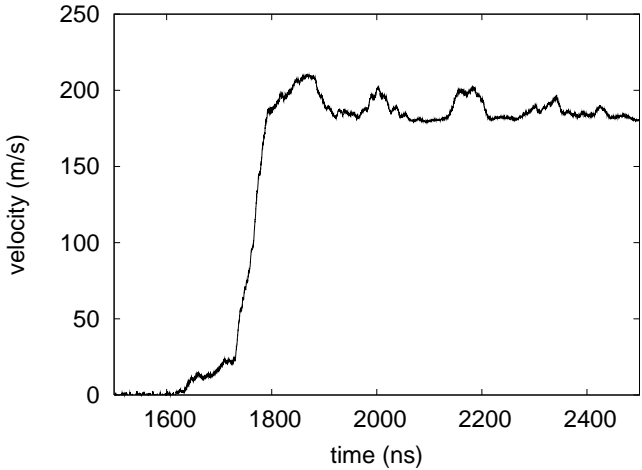}
\end{center}

\caption{Example velocity history at the free surface of the sample in 
   a baseplate experiment (shot 14136).}
\label{fig:14136_vpoint}
\end{figure}

In the release window and baseplate experiments, the velocity history at the
surface of the NiAl generally exhibited a precursor to a particle speed
of 15-30\,m/s ahead of the main shock wave.
This precursor was presumably an elastic wave.
The rising part of the shock generally exhibited some structure, consistent
with reverberations of the elastic wave between the surface and the approaching
shock.
The peak velocity was followed by deceleration of 25-30\,m/s
(with some outliers) and reverberations, consistent with spallation and ringing.
(Figs~\ref{fig:14128_vcmp} to \ref{fig:14136_vpoint}.)

\begin{figure}
\begin{center}
\includegraphics[scale=0.72]{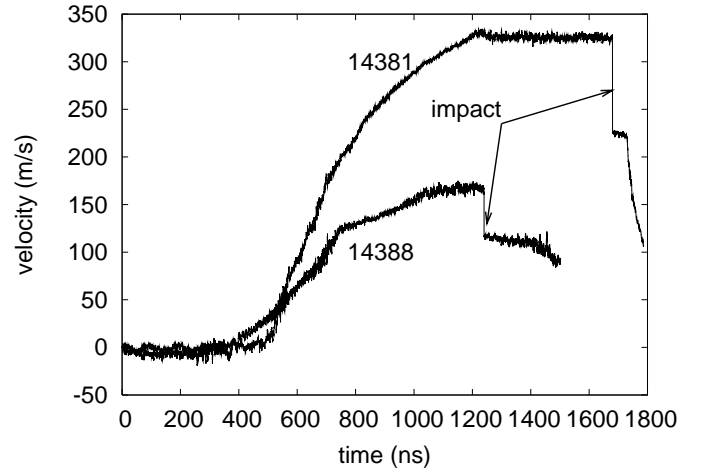}
\end{center}

\caption{Example velocity histories from window impact experiments.}
\label{fig:winimpactcmp}
\end{figure}

In the window impact experiments, 
the flyer speed just before and just after impact
were determined from the VISAR record
(Fig.~\ref{fig:winimpactcmp}).

\subsection{Analysis}
\begin{figure}
\begin{center}
\includegraphics[scale=0.72]{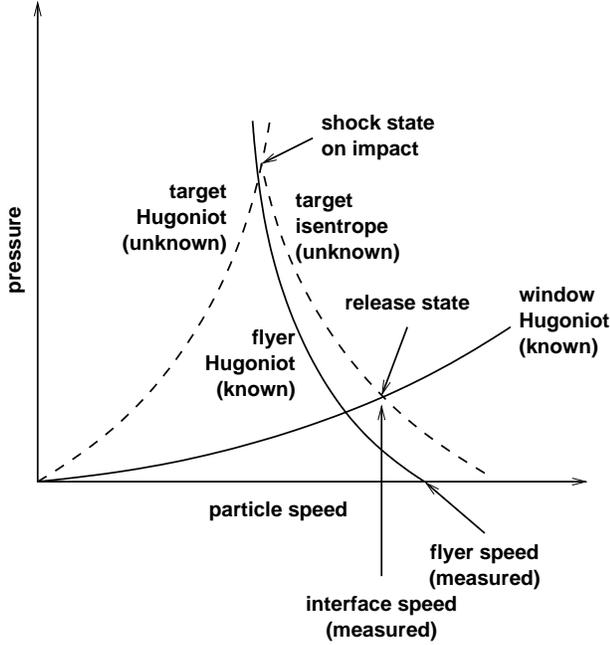}
\end{center}

\caption{Schematic of construction used to deduce Hugoniot point
   from window release data with a flyer of known material.}
\label{fig:asymreleaseschem}
\end{figure}

\begin{figure}
\begin{center}
\includegraphics[scale=0.72]{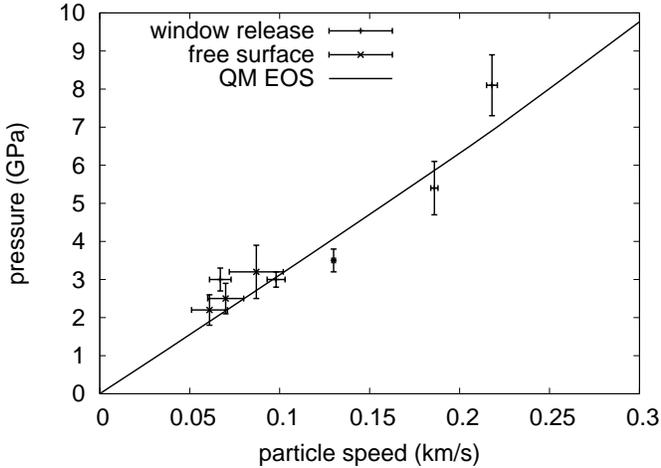}
\end{center}

\caption{NiAl Hugoniot points deduced from the impact of Cu flyers,
   compared with the {\it ab initio} equation of state.}
\label{fig:upp}
\end{figure}

The impact of a Cu flyer onto a NiAl target, releasing into a window,
allowed a point on the principal Hugoniot to be deduced,
relying on the assumption that the NiAl release isentrope through the
shock state generated by the impact was equal to the principal Hugoniot
(Fig.~\ref{fig:asymreleaseschem}).
At the pressures of a few gigapascals generated in these experiments, 
this assumption is accurate to a percent or better.
This analysis is simplest for release into vacuum,
where the particle speed in the sample can be estimated as half of the
free surface speed.
The difference between this and the flyer speed is equal to the 
particle speed in the flyer, and the Hugoniot of the flyer material
can be used to find the shock pressure.
Reference Hugoniots were calculated from published EOS \cite{Steinberg96}.
Hugoniot states were obtained, with pressures between 2 and 8\,GPa.
States deduced by window release and free surface velocity were consistent.
The principal Hugoniot from the {\it ab fere initio} EOS passed as closely
through the data as any unique line is likely to.
(Fig.~\ref{fig:upp}.)

A complication of the window impact experiments 
compared with the Cu flyer experiments 
was that the stress state at the interface between the
sample and the window depends more sensitively on the elastic strain in both
components.
While the elastic constants may be known,
the flow stress (or yield strength) is not known for all time scales.
This was not such a concern for the Cu flyer experiments because the flow
stress of Cu and PMMA are much lower than for LiF \cite{Steinberg96}.
Calculations were made using different assumptions about the flow stress,
to assess the uncertainty and the likely behavior of LiF in these experiments.

\begin{figure}
\begin{center}
\includegraphics[scale=0.72]{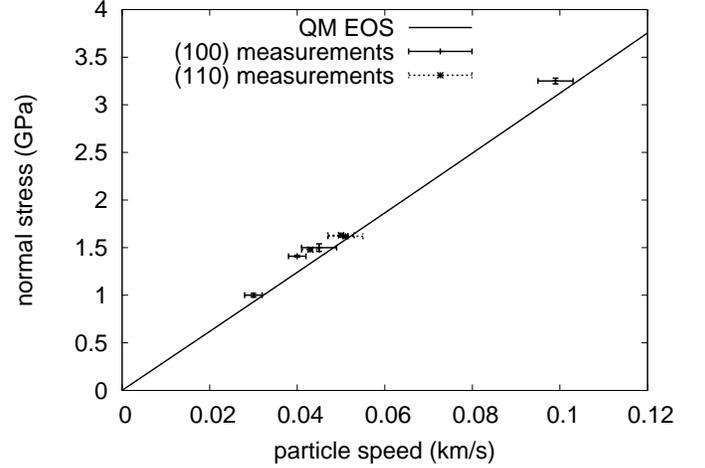}
\end{center}

\caption{Hugoniot points deduced for nickel -- aluminum alloy,
   compared with the {\it ab initio} equations of state.}
\label{fig:hydroupp}
\end{figure}

Neglecting elasticity,
a measurement of the particle speed immediately before
and after impact ($u_0$ and $u_1$) can be used
to determine a state on the principal Hugoniot of the flyer
with reference to the principal Hugoniot of the window,
which must be known.
Immediately after impact -- for a time dictated mainly by the shock
and release transit time through the flyer and window -- 
the material in both components at the impact surface is at the
same pressure and traveling at the same speed.
The Hugoniots are expressed in pressure -- particle speed ($p$ -- $u_p$)
space.
In the shocked region, the particle velocity with respect
to the undisturbed material in the window is $u_1$,
and in the flyer $u_0-u_1$.
As the same pressure exists in both materials, and it can be calculated from 
the Hugoniot of the window material $p_W(u_1)$, 
a Hugoniot state can be determined in the flyer: $(u_0-u_1,p_W(u_1))$.
The Rankine-Hugoniot relations can then be used to calculate
the other state parameters: mass density $\rho$, shock speed $u_s$,
and the change in specific internal energy $e$
(Table~\ref{tab:uppmodels} and Fig.~\ref{fig:hydroupp}).

More rigorously, the materials at the impact interface are at the same
state of normal stress $s_{11}$ rather than pressure, and
under uniaxial compression.  By the same arguments as above, 
if the shock Hugoniot of the window -- now considering
its response as a stress tensor $s$, at least the normal component -- is known, 
then the normal stress of the flyer material is thus also known.
The stress components for a solid depend on the elastic strain and
the elastic constants.
Elastic strain can be relieved by plastic flow, which limits the
deviation of normal stress from the hydrostatic pressure.
The elastic constants of common window materials are well-known,
but the finite rate of plastic flow -- which can also be considered in terms of
a flow stress or yield stress which varies with strain rate -- means that
experiments exploring the response on different time scales may induce
significantly different degrees of elastic strain for the same compression.
Our experiments used flyers which were relatively thin compared with
previous experiments, so it is possible that the elastic strains were
somewhat greater.

Another aspect of this complication is that we ideally want to
determine the EOS.  Data from window impact experiments comprise the EOS
plus the elastic shear stress.
For a single experimental measurement in isolation, there is no way
to separate the hydrostatic pressure (EOS) from the elastic stress,
but some values can be extracted given data from several experiments
at different pressure, and/or knowledge of the EOS or elastic and plastic
behavior.
In the present case, relevant complementary data include the STP 
elastic constants of NiAl, our QM EOS,
and low strain-rate measurements of the flow stress.
We considered the limiting case of no plastic flow (i.e. maximum
elastic stress for a given uniaxial compression),
and also cases with varying flow stresses in the window and the NiAl flyer.

For predicted shock Hugoniots including elasticity, 
the elastic stress was added to the Hugoniot pressure calculated from the EOS.
For each state on the Hugoniot, the uniaxial compression 
$\eta=\rho/\rho_0$ was used to calculate the strain tensor
(using the Green-St~Venant finite strain measure \cite{Fung69})
\begin{equation}
e=\left(
\begin{array}{ccc}
\frac 12\left(1/\eta^2-1\right) & 0 & 0 \\
0 & 0 & 0 \\
0 & 0 & 0
\end{array}
\right)
\end{equation}
and hence the strain deviator
\begin{equation}
\epsilon=\left(1-1/\eta^2\right)\left(
\begin{array}{ccc}
\frac 23 & 0 & 0 \\
0 & \frac 16 & 0 \\
0 & 0 & \frac 16
\end{array}
\right).
\end{equation}
The corresponding elastic stress is given by
\begin{equation}
\pdiffl{[s]_{ij}}{[e]_{kl}} = [c(\rho)]_{ijkl}
\end{equation}
where $c$ is the tensor of elastic constants, strictly a function
of $e$ too.
In deviatoric form,
\begin{equation}
s\equiv\sigma(\rho,\epsilon)-p(\rho)I
\end{equation}
where $p(\rho)$ is the EOS and
\begin{equation}
\pdiffl{[\sigma]_{ij}}{[\epsilon]_{kl}} = [c(\rho)]_{ijkl}.
\end{equation}
Plastic flow acts to limit the components of the elastic strain deviator 
$\epsilon$;
in common practice it may be expressed as a constraint on the magnitude 
of the stress deviator.

The EOS is a more general representation of the bulk modulus 
\begin{equation}
B = \frac 13\left(c_{11} + 2 c_{12}\right).
\end{equation}
Comparing the response of a crystal to uniaxial compression along $[100]$,
with stiffness to uniaxial compression $c_{11}$,
we can define a shear modulus 
\begin{equation}
\mu = \frac 12\left(c_{11} - c_{12}\right).
\end{equation}
Using the published STP elastic constants for NiAl 
($c_{11} = 211.5$\,GPa, $c_{12} = 143.2$\,GPa, $c_{44} = 112.1$\,GPa)
\cite{Wasilewski66}, $\mu=34.2$\,GPa.

\begin{figure}
\begin{center}
\includegraphics[scale=0.72]{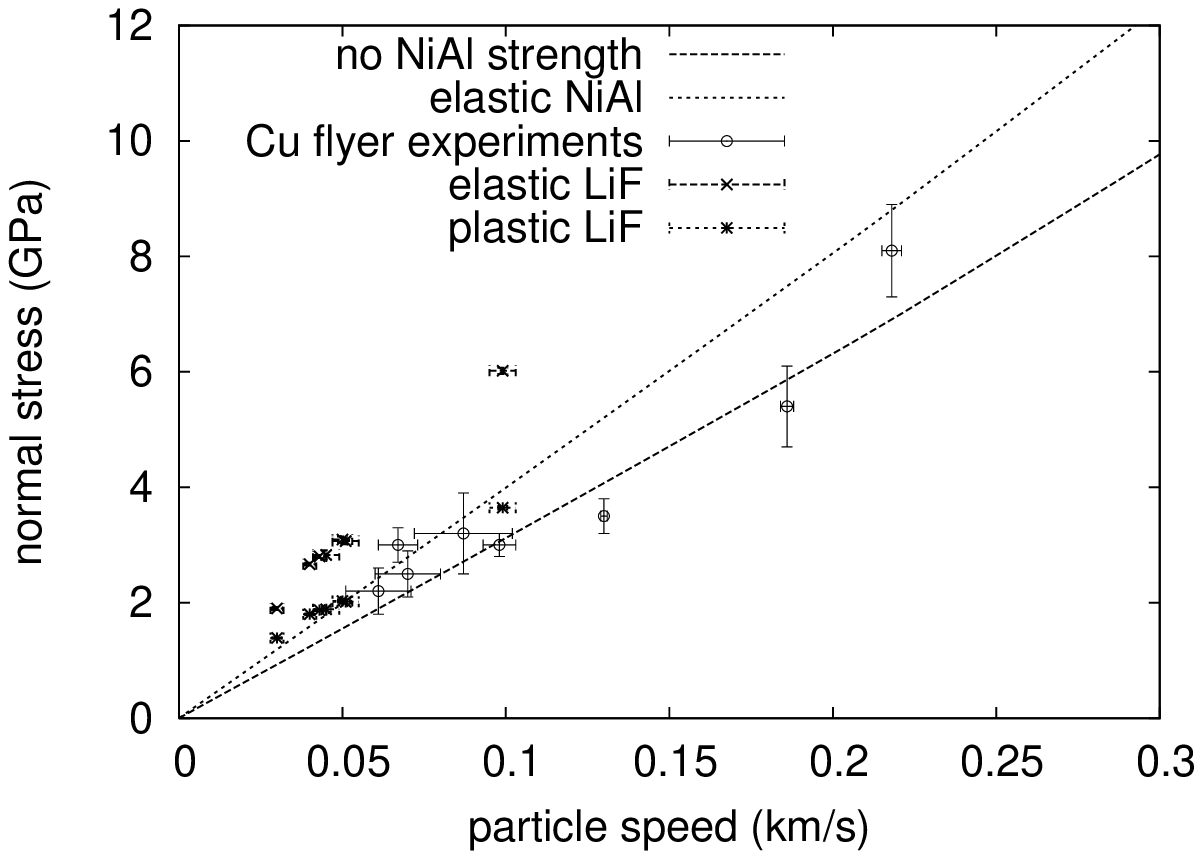}
\end{center}

\caption{Hugoniot points deduced for NiAl,
   compared with the {\it ab initio} equation of state.
   Points with large error bars are from the Cu flyer experiments.}
\label{fig:cmpupp_ep}
\end{figure}

\begin{figure}
\begin{center}
\includegraphics[scale=0.72]{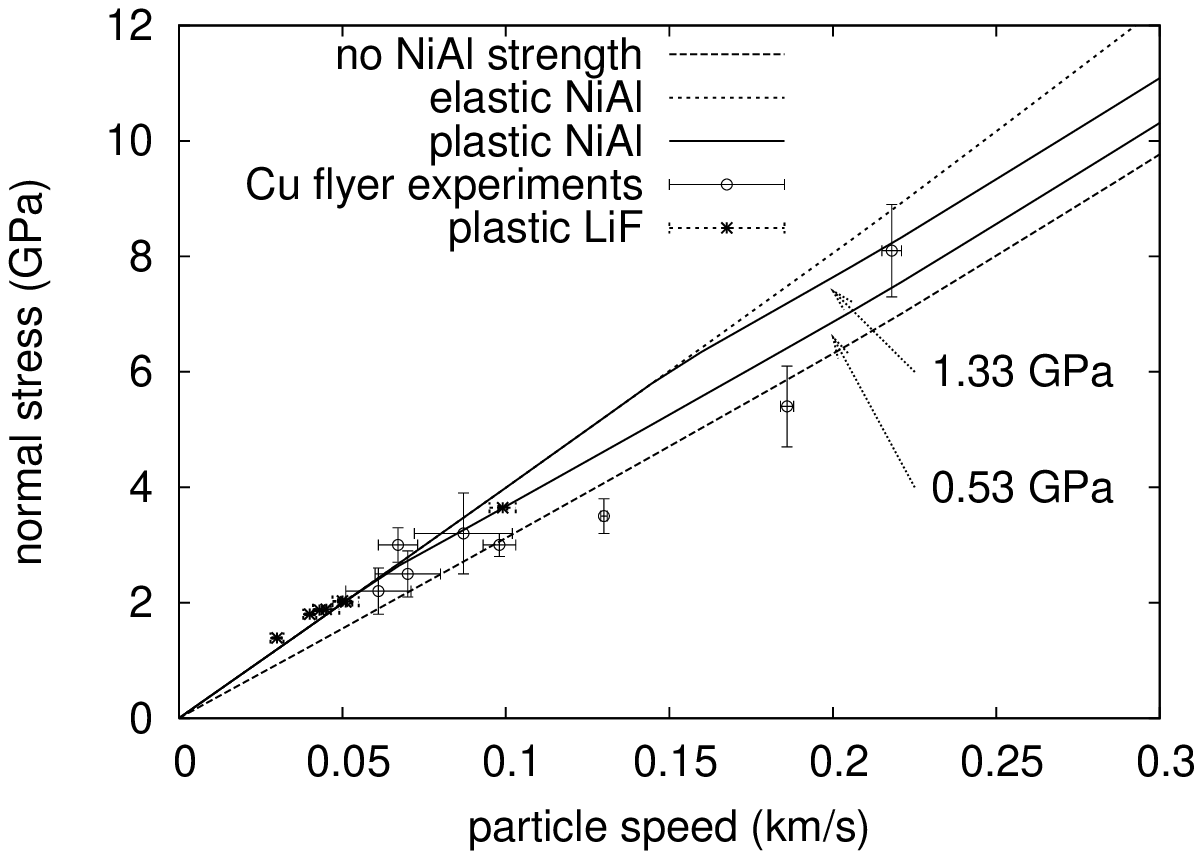}
\end{center}

\caption{Hugoniot points deduced for NiAl,
   compared with the {\it ab initio} equation of state.
   Points with large error bars are from the Cu flyer experiments.}
\label{fig:cmpupp_ep1}
\end{figure}

The shear modulus was used to calculate the deviatoric correction to
the shock Hugoniot from the EOS, for uniaxial compression.
These corrections were calculated for purely elastic response,
and also for different values of the flow stress $Y$ chosen to
improve the match to the experimentally-measured shock states
(Table~\ref{tab:uppmodels}).
If the LiF window was treated as purely elastic then no treatment of the NiAl
reproduced the experimental states.
If the LiF was treated as elastic-plastic with flow stress $Y=0.36$\,GPa
as deduced from gas gun experiments, then the measured states
were reproduced fairly well with a flow stress of 0.53\,GPa in the NiAl.
(Figs~\ref{fig:cmpupp_ep} and \ref{fig:cmpupp_ep1},
also showing shock states deduced from the Cu flyer
experiments.
The Cu flyer data were not corrected
for elastic response in the window: the correction is smaller for PMMA.)

\begin{table*}
\caption{Shock Hugoniot states deduced from window impact experiments,
   using different constitutive models for LiF and NiAl.}
\label{tab:uppmodels}
\begin{center}
\begin{tabular}{|c|rcl|rcl|rcl|rcl|rcl|}\hline
{\bf shot} & \multicolumn{6}{|c|}{\bf particle speed (m/s)}
 & \multicolumn{9}{|c|}{\bf normal stress (GPa)} \\
{\bf shot} & \multicolumn{6}{|c|}{\bf for different LiF models}
 & \multicolumn{9}{|c|}{\bf for different NiAl models} \\ \cline{2-16}
 & \multicolumn{3}{|c|}{\bf hydrodynamic}
 & \multicolumn{3}{|c|}{\bf $Y=0.36$\,GPa}
 & \multicolumn{3}{|c|}{\bf hydrodynamic}
 & \multicolumn{3}{|c|}{\bf elastic}
 & \multicolumn{3}{|c|}{\bf $Y=0.53$\,GPa} \\
\hline
{\bf (100)} & & & & & & & & & & & & & & & \\
14381 & 99 & $\pm$ & 4 & 226 & $\pm$ & 2
   & 3.25 & $\pm$ & 0.03 & 6.02 & $\pm$ & 0.05 & 3.64 & $\pm$ & 0.03 \\
14389 & 30 & $\pm$ & 2 &  72 & $\pm$ & 1
   & 1.00 & $\pm$ & 0.02 & 1.90 & $\pm$ & 0.03 & 1.39 & $\pm$ & 0.01 \\
14390 & 45 & $\pm$ & 4 & 107 & $\pm$ & 3
   & 1.50 & $\pm$ & 0.04 & 2.83 & $\pm$ & 0.08 & 1.89 & $\pm$ & 0.04 \\
14406 & 40 & $\pm$ & 2 & 101 & $\pm$ & 1
   & 1.41 & $\pm$ & 0.01 & 2.67 & $\pm$ & 0.03 & 1.80 & $\pm$ & 0.01 \\
\hline
{\bf (110)} & & & & & & & & & & & & & & & \\
14387 & 50 & $\pm$ & 3 & 117 & $\pm$ & 2
   & 1.63 & $\pm$ & 0.02 & 3.09 & $\pm$ & 0.05 & 2.03 & $\pm$ & 0.03 \\
14388 & 51 & $\pm$ & 4 & 116 & $\pm$ & 2
   & 1.62 & $\pm$ & 0.02 & 3.07 & $\pm$ & 0.05 & 2.01 & $\pm$ & 0.03 \\
14405 & 43 & $\pm$ & 2 & 106 & $\pm$ & 1
   & 1.48 & $\pm$ & 0.02 & 2.80 & $\pm$ & 0.03 & 1.87 & $\pm$ & 0.01 \\
\hline\end{tabular}
\end{center}

{\small
The hydrodynamic model for NiAl was used with the hydrodynamic model for LiF.
The elastic and elastic-plastic models for NiAl were used with the
elastic-plastic model for LiF.
}
\end{table*}

\begin{figure}
\begin{center}
\includegraphics[scale=0.72]{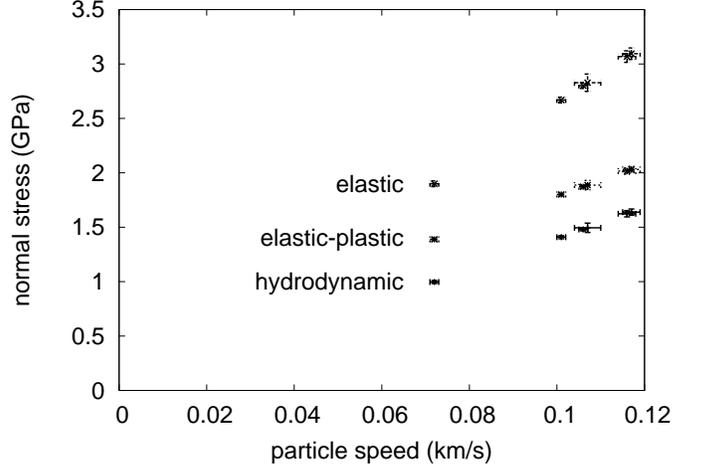}
\end{center}

\caption{Stress states deduced for NiAl,
   using different assumptions about plasticity in the LiF window.}
\label{fig:stressupp}
\end{figure}

The precursor waves and the deceleration following the peak of the shock
can be used to infer flow stresses and spall strengths.
However, simple models parameterized from these data can be misleading,
so the detailed constitutive behavior of NiAl will be reported
separately in the context of microstructural response.
The true (i.e. normal) stress to induce plastic flow at lower strain rates
has been reported at around 1.5\,GPa
\cite{Maloy95}, corresponding to a flow stress of around 0.4\,GPa.
One would expect the stresses to be
at least as large on the shorter time scales of the laser experiments,
so the flow stress inferred from the window impact experiments is quite
plausible.

\section{Conclusions}
The frozen-ion compression curve for NiAl in the CsCl structure was predicted 
using {\it ab initio} quantum mechanical calculations of the electron band
structure.
The calculations reproduced the lattice parameter at $p=0$
to $\sim$1\%\ with no corrections, equivalent to a discrepancy $\sim 7$\,GPa.
The {\it ab initio} pressure-volume relation calculated using the
Hellmann-Feynman theorem was not perfectly consistent with the {\it ab initio}
energy-volume relation;
this reflects the lower precision of the stress calculations.
The Rose functional form, found to fit the energy-volume relation for
compression of a wide range of elements, was found to fit the energy-volume
relation for compression of NiAl, but deviated significantly in expansion.
The electron band structure was used to predict electron-thermal excitations.
The mechanical response in the equation of state and shock Hugoniot 
did not alter much when the electron-thermal contribution was included, though
the Hugoniot pressure-temperature relation varied by
around 1\%\ per thousand kelvin.
The charge distribution in the electron ground states was used to predict 
{\it ab initio} phonon modes, from the forces on the atoms when one was
displaced from equilibrium.
The phonon density of states was quite sensitive to the magnitude of the
displacement -- though the restoring force on the displaced atom itself
suggested that the effective potential it experienced was essentially
harmonic -- but thermodynamically-complete equations of state constructed using 
the phonon modes were not sensitive to the details of the density of phonon
states up to $\sim$100\,GPa.
The quasiharmonic equations of state reproduced published measurements
of isothermal compression of NiAl extremely well, and were also consistent
with measurements of the shock compression.

Laser-driven flyer experiments were performed to measure states on the
principal shock Hugoniot of NiAl, by impacting Cu flyers into NiAl targets
and by impacting NiAl flyers against LiF windows.
Shock transit times were not measured accurately enough to constrain the EOS.
Shock states from the Cu flyer data were obtained with respect to the 
Hugoniot of Cu or PMMA.
These states were consistent with the {\it ab initio} EOS for NiAl.
Detailed interpretation of the NiAl flyer data depends on the elastic-plastic
behavior of the NiAl and the LiF.
Ignoring these elastic contributions, the shock states were consistent
with theoretical EOS.
If the LiF was assumed to response elastically, the stress states deduced
were implausibly high.
If it was assumed to response with the same flow stress as observed in
gas gun experiments -- which typically explore somewhat longer time scales
-- then it was possible to reproduce the shock states by taking either
EOS and adjusting the flow stress in the NiAl.
Reasonable agreement was obtained for the quasiharmonic EOS 
with a flow stress of 0.53\,GPa,
which is consistent with the range of values deduced from the amplitude of the
elastic precursor wave in other shock loading experiments.
The quasiharmonic EOS -- which was found to reproduce quasistatic compression
data extremely well -- is thus consistent with the shock measurements.

\section*{Acknowledgments}
John Brooks and Darrin Byler helped prepare the NiAl samples.
The TRIDENT staff including Randy Johnson, 
Tom Hurry, Tom Ortiz, Fred Archuleta, Nathan Okamoto, and Ray Gonzales
were crucial to the conduct of the experiments.
Sheng-Nian Luo gave useful advice on the analysis of the experimental data,
and David Schiferl on the accuracy of diamond-anvil cell measurements.
The computer program CASTEP was made available courtesy of Accelrys and the U.K.
Car-Parrinello Consortium.
Function fitting was performed using
Wolfram Research Inc.'s computer program `Mathematica' (V5.0)
and Wessex Scientific and Technical Services Ltd's
C++ mathematical subroutine library (V2.1).
We would like to thank the manuscript reviewer for helpful observations
and suggestions.
This work was performed under the auspices of the U.S. Department of Energy
under contract W-7405-ENG-36, 
as part of the Laboratory-Directed Research and Development (Directed Research)
project on `Shock Propagation at the Mesoscale'
(2001-4; principal investigator: Aaron Koskelo).

\end{document}